**The molecular mechanism of a *cis*-regulatory adaptation in yeast**

Jessica Chang[1], Yiqi Zhou[1], Xiaoli Hu[1,2], Lucia Lam[3,4], Cameron Henry[1], Erin M. Green[1], Ryosuke Kita[1], Michael S. Kobor[3,4], and Hunter B. Fraser[1]*

[1]Department of Biology, Stanford University, Stanford CA 94305.
[2]Present address: Key Laboratory of Marine Genetics and Breeding, Ministry of Education, College of Marine Life Sciences, Ocean University of China, Qingdao, China.
[3]Department of Medical Genetics, University of British Columbia, Vancouver, British Columbia, Canada

[4]Centre for Molecular Medicine and Therapeutics, Child and Family Research Institute, Vancouver, British Columbia V5Z 4H4, Canada

*Correspondence: hbfraser@stanford.edu

**Abstract**

Despite recent advances in our ability to detect adaptive evolution involving the *cis*-regulation of gene expression, our knowledge of the molecular mechanisms underlying these adaptations has lagged far behind. Across all model organisms the causal mutations have been discovered for only a handful of gene expression adaptations, and even for these, mechanistic details (e.g. the *trans*-regulatory factors involved) have not been determined. We previously reported a polygenic gene expression adaptation involving down-regulation of the ergosterol biosynthesis pathway in the budding yeast *Saccharomyces cerevisiae.* Here we investigate the molecular mechanism of a *cis*-acting mutation affecting a member of this pathway, *ERG28*. We show that the causal mutation is a two-base deletion in the promoter of *ERG28* that strongly reduces the binding of two transcription factors, Sok2 and Mot3, thus abolishing their regulation of *ERG28*. This down-regulation increases resistance to a widely used antifungal drug targeting ergosterol, similar to mutations disrupting this pathway in clinical yeast isolates. The identification of the causal genetic variant revealed that the selection likely occurred after the deletion was already present at high frequency in the population, rather than when it was a new mutation. These results provide a detailed view of the molecular mechanism of a *cis*-regulatory adaptation, and underscore the importance of this view to our understanding of evolution at the molecular level.

**Author Summary**

Evolutionary adaptation is the process that has given rise to the ubiquitous, yet remarkable, fit between all living organisms and their environments. The molecular mechanisms of these adaptations have been a subject of great interest, but we still know very little about their mechanisms, particularly in the case of regulatory adaptations. In this work, we investigate the molecular mechanism of a regulatory adaptation that we previously identified in *ERG28*, a component of the ergosterol biosynthesis pathway in budding yeast. Ergosterol is an abundant lipid component of the fungal plasma membrane, and is of major biomedical importance, being targeted by numerous antifungal drugs. We identified the causal mutation underlying the *ERG28* adaptation, a two-base deletion in its promoter which leads to lower abundance of its mRNA. This deletion acts via disrupting the binding of at least two transcription factors, Mot3 and Sok2, to the promoter. The deletion increases resistance to a widely used antifungal drug, Amphotericin B, which targets ergosterol. This effect is reminiscent of misregulation of the ergosterol pathway observed in clinical yeast isolates that have evolved resistance to Amphotericin B. Our results may therefore have medical implications, while also advancing our basic understanding of evolutionary mechanisms.

## Introduction

Evolutionary adaptation is the process that has given rise to the ubiquitous, yet remarkable, fit between all living organisms and their environments [1]. The origins of these adaptations at the molecular level have been a subject of great interest, with active debate surrounding the relative roles of two major classes of molecular mechanism: changes in protein sequences *vs*. changes in the expression levels/patterns of those proteins [2-5]. Until recently, the evidence cited in favor of both mechanisms was either anecdotal (involving studies of single genes) or theoretical in nature [2-4]. However, the advent of methods for characterizing gene expression adaptation genome-wide [6-9] (as well as methods for measuring *cis*-regulatory changes that may or may not be adaptive [10-11]) has paved the way for this question to be addressed in an unbiased, systematic fashion [5].

Although the distinction between protein sequence *vs*. gene expression regulation is important, it is only one of many levels at which molecular mechanisms can be distinguished. For example among *cis*-regulatory adaptations, mutations might act via alterations in transcription factor (TF) binding, nucleosome positioning, mRNA processing, binding of RNA-binding proteins, etc. As the field matures, it is likely that the distinctions between these more detailed mechanistic levels will be of increasingly greater interest, since only by investigating these mechanisms will we fully understand the nature of adaptation at the molecular level.

In order to investigate the molecular mechanism of an adaptation, it is generally necessary to first identify the causal mutation(s) (though see [12]). This prerequisite has been a significant bottleneck in studies of *cis*-regulatory adaptation. Because we cannot computationally predict the effects of most non-coding mutations, and such mutations can act at long distances from their target genes in many species (resulting in a large search space), only a handful of causal mutations underlying *cis*-regulatory adaptations have been reported. For example, large deletions of an enhancer driving the pelvic expression of the *Pitx1* gene in sticklebacks have been found to result in adaptive pelvic reduction in freshwater populations [13]. In another case, five non-coding mutations at the *ebony* locus contributed to dark abdominal pigmentation found in high-altitude populations of *Drosophila melanogaster* [14] (although other examples exist where causal *cis*-regulatory mutations have been identified in various model organisms [15-17], these have not been shown to be adaptive). However even for these intensively studied *cis*- regulatory adaptations, the molecular mechanisms by which the causal mutations act—e.g. which TFs and/or epigenetic states are affected by the mutations—remain unknown.

We previously reported a genome-wide scan for gene expression adaptation between two strains of the budding yeast *Saccharomyces cerevisiae*: a laboratory strain (BY4716, hereafter "BY") and a vineyard strain (RM11-1a, hereafter "RM") [8]. We found that over 200 genes had likely been subject to recent positive selection in these strains via the action of reinforcing *cis* and *trans*-acting regulatory adaptations. Among these genes, there was a strong enrichment of down-regulating mutations in one metabolic pathway: ergosterol biosynthesis. Ergosterol is an abundant lipid component of the fungal plasma membrane, and is of major biomedical importance, being targeted by numerous antifungal drugs [18]. Indeed, a common mechanism of resistance to ergosterol-targeting drugs (such as amphotericin B) is reducing ergosterol levels via disruption of this pathway [18-21]. We previously found that six genes within the pathway (underlined and red in Figure 1A) showed the strongest signs of selection, based on patterns of reinforcing *cis/trans*-regulatory mutations, as well as a population-genetic signature of selective sweeps in the genomes of multiple strains [8]. This represents the first known example of a polygenic gene expression adaptation, from any species. Here, we sought to gain a deeper understanding of this adaptation.

## Results

### *Further characterization of the ergosterol pathway adaptation*

Because our initial identification of the polygenic gene expression adaptation within the ergosterol (ERG) biosynthesis pathway was based on expression data from genome-wide microarrays [22], we first sought to more precisely measure the *cis*-regulatory divergence at these loci. This divergence can be measured for any gene as the ratio of mRNA abundances of the two alleles present in a hybrid diploid: in the absence of *cis*-acting differences, the mRNA from the two alleles will be present in equal amounts (as they are in the genomic DNA), whereas they will be unequal in the presence of *cis*-regulatory divergence. To measure this ratio we employed pyrosequencing, a method that accurately quantifies allelic ratios at individual heterozygous sites [23].

Of the six genes we previously implicated, five were amenable to this approach (the sixth, *ERG26*, lacked any BY/RM sequence differences in its mRNA, so the alleles could not be distinguished). All five showed reproducible allelic imbalance in the expected direction (lower expression from the BY allele), with magnitudes ranging from 1.13-fold to 1.94-fold (Figure 1B). This result confirms that the "local eQTL" (genetic markers showing a statistical association with a nearby gene's expression level) previously mapped for these genes [22] likely represent *cis*-acting genetic variants.

To investigate if the polygenic adaptation extends beyond the six genes we originally identified, we also performed pyrosequencing on three additional ERG genes adjacent in the pathway to those already implicated: *ERG25*, *ERG27*, and *ERG2* (allelic bias of *ERG1*, the other adjacent pathway member, could not be measured because it has no sequence differences between BY and RM). We found reproducible allelic bias in favor of RM for both *ERG25* and *ERG27*, but not for *ERG2* (Figure 1B). This suggests that the adaptive down-regulation extends to a total of at least eight genes, forming a contiguous block within the ERG pathway (Figure 1A, in red) that has been specifically targeted by natural selection.

Interestingly, in addition to the clear clustering of the down-regulated genes within the pathway, the genes with the strongest *cis*-regulatory differences correspond precisely to the core proteins in a stable complex organized by Erg28. Erg28 is the only known member of the ERG pathway lacking enzymatic activity; it is an endoplasmic reticulum transmembrane protein, highly conserved across eukaryotes (including humans), that acts as a scaffold promoting co-localization of ERG enzymes [24-26]. Erg28 physically interacts most strongly with Erg27 (and is thus shown next to Erg27 in Figure 1), but has also been found to interact strongly with itself and three other proteins: Erg25, Erg6, and Erg11; its other interactions are significantly weaker [24]. These five interacting proteins are not only all components of the polygenic adaptation (Figure 1), but are specifically those components with the strongest *cis*-acting down-regulation: all five have at least 1.25-fold differences between RM and BY alleles, while no other genes quite reach this threshold (Figure 1B). This pattern suggests that the precise magnitude of down-regulation may be influenced both by pathway position and by membership in the protein complex organized by Erg28.

*Pinpointing a causal adaptive mutation*

We decided to focus on *ERG28* for further investigation. Not only is Erg28 the central member of the protein complex apparently targeted by natural selection, but sequence divergence in its promoter region was also minimal: there are only two sequence differences between BY and RM in the 590 bp upstream of the *ERG28* transcription start site (TSS). These are one two-bp deletion (located in an 11 bp poly-A tract 112 bp upstream of the TSS, termed the AA112Δ allele), and one T/C SNP (229 bp upstream of the TSS, the T229C allele) (Figure 2A). Because promoters in *S. cerevisiae* are compact (generally < 400 bp [27]), we decided to focus on these two candidate variants.

To definitively identify the mutation(s) underlying a *cis*-regulatory adaptation, the mutations must be individually tested for their effects on expression of the associated gene. Therefore we constructed allelic replacement strains in which individual BY variants were introduced into the RM genome. Using a method of *in vivo* site-directed mutagenesis known as *delitto perfetto* [28], we engineered strains that differed only by the desired mutation. We refer to the two resulting strains as RM AA112Δ and RM T229C (Figure 2b).

If a mutation can fully account for the 1.30-fold *cis*-acting difference between the RM/BY alleles of *ERG28* (Figure 1B), and no additional mutations have any effect, then this mutation can be deemed causal. To test if this was the case for either of our candidate mutations, we measured the expression level of *ERG28* in each strain, as well as in wild-type RM, by quantitative PCR (qPCR). While we found no effect of the T229C mutation (1.05-fold difference), we observed that the AA112Δ mutation led to a 1.26-fold decrease in mRNA level (Figure 2c), indistinguishable from the 1.30-fold change expected for the causal mutation(s).

To further test if the AA112Δ mutation could fully account for the RM/BY difference, we mated the RM AA112Δ strain with BY, and measured the allelic ratio of *ERG28* mRNA in the resulting diploid strain. The causal mutation would be expected to reduce the 1.30-fold allelic difference to ~1, while any non-causal mutation would have the same the 1.30-fold allelic imbalance found in the BY/RM hybrid. Consistent with the qPCR results, the RM AA112Δ/BY hybrid strain showed a 1.03-fold difference between alleles, while the RM T229C/BY hybrid showed a 1.27-fold difference (Figure 2d). Together, these results suggested that the AA112Δ mutation likely accounted for all, or nearly all, of the *cis*-acting divergence at *ERG28* between RM and BY.

*The molecular mechanism of the* ERG28 cis-*regulatory adaptation*

We considered two potential mechanisms for how the AA112Δ mutation may be down-regulating transcription: nucleosome positioning and TF binding. Both processes are known to play important roles in determining rates of transcription initiation, and could potentially be affected by a 2-bp deletion.

Nucleosome positioning was an especially plausible mechanism because the 11-bp poly-A sequence in which the 2-bp deletion occurred is a strong nucleosome-disfavoring sequence [29]. Therefore we took advantage of published data on genome-wide nucleosome positions from BY and RM [30] to determine whether the nucleosome overlapping the deletion was affected. There was no significant difference between BY and RM in the nucleosomal occupancy or positioning at this location (nor was it differentially acetylated on histone H3 lysine 14 [30]), suggesting that nucleosome occupancy was not greatly affected by this deletion.

We therefore turned to TF binding as a second possible mechanism. Utilizing a published map of putative TF binding sites [31] we identified two highly conserved (across *Saccharomyces sensu stricto*) binding sites for the TFs Mot3 and Sok2, flanking the deletion (Figure 3a). Mot3 is a well-known repressor of ERG pathway genes, exerting its greatest effect in hypoxic or hyper-osmotic conditions [32-33], whereas Sok2 has not been previously linked to the ERG pathway to our knowledge. Neither binding site motif is directly affected by the 2-bp deletion; rather the only effect is on their spacing, reducing the distance between motif centers from 16 bp to 14.

To test if the AA112Δ deletion may affect the regulation of *ERG28* by either of these two TFs, we created knockout strains for each TF in both the wild-type RM and RM AA112Δ backgrounds. Several outcomes are possible (Figure 3b). First, if the TF does not regulate *ERG28*, then deleting it should have no effect in either genetic background. Second, if the TF does regulate *ERG28* but is not affected by the AA112Δ deletion, then the effect of TF deletion should be equal in the two backgrounds. Finally, if the AA112Δ deletion is affecting the TF's regulation of *ERG28,* then the effect of TF deletion will depend on the background—for example, having an effect on *ERG28* expression in wild-type RM but not in RM AA112Δ.

Consistent with Mot3's known role as a repressor of ERG pathway genes, we found that *ERG28* was induced 1.85-fold in an RM *mot3*Δ strain compared with wild-type RM (Figure 3c).

Likewise, Sok2 was found to be an activator of *ERG28*, with 1.21-fold lower expression in RM *sok2*Δ compared to wild-type RM. However neither TF had any measurable effect on *ERG28* expression when deleted from the RM AA112Δ strain (Figure 3c). This suggests that although both TFs regulate *ERG28* in RM, this regulation was abolished by the 2-bp deletion.

The effect of AA112Δ on regulation of *ERG28* by Mot3 and Sok2 suggested that their binding to the promoter may be affected by the deletion. To investigate this, we performed chromatin immunoprecipitation (ChIP). Specifically, we HA-tagged both TFs in both wild-type RM and RM AA112Δ backgrounds, and quantified their binding to specific regions by quantitative PCR (qPCR). We found that for both factors, binding at the *ERG28* promoter was reduced in RM AA112Δ, compared to wild-type RM: Sok2 showed ~19-fold lower binding, while Mot3 had ~31-fold lower binding (Figure 3d). This suggests that the loss of *ERG28* regulation by these TFs in the AA112Δ background (Figure 3c) is likely due to their severely reduced binding.

### *Fitness effects of the* ERG28 cis-*regulatory adaptation*

In order to investigate the phenotypic effects of the AA112Δ allele, we measured the growth rates of our engineered strains and RM in several environments (see Materials and Methods). While we did not observe any fitness advantage of the RM AA112Δ strain in most conditions (e.g. rich synthetic defined [SD] media; paired t-test $p = 0.46$ for RM AA112Δ *vs*. RM and $p = 0.83$ for RM T229C *vs*. RM; Figure 4a), we did find a growth advantage of this strain in the presence of the antifungal drug amphotericin B (Figure 4b). Specifically, RM AA112Δ had a 1.3% higher growth rate than RM when grown in the presence of the drug ($p = 0.014$), whereas RM T229C had no measurable difference from RM ($p = 0.86$). This suggests that the fitness benefit conferred by the AA112Δ allele is condition-specific.

### *Insights into the selection on* ERG28 cis-*regulation*

Our identification of the AA112Δ allele as causal allows us to examine the distribution of this adaptive mutation across other yeast strains, in order to study its history. In particular, we wished to address the question of whether the selection occurred when the deletion was a new mutation that just recently arose (e.g. in the laboratory), or whether it was present as "standing variation" in *S. cerevisiae* for some time before the selection occurred. Population geneticists have theorized about the consequences of selection acting on pre-existing variation, as opposed to waiting for rare advantageous mutations to occur, but few clear examples exist [34-36].

To distinguish between these alternatives, we first examined the distribution of the AA112Δ allele across a set of 36 sequenced strains of *S. cerevisiae* [37]. The deletion is present in 12/36 sequenced strains (in addition to BY; Figure S1). These 12 strains are diverse in terms of both geography (from the Americas, Asia, Africa, and Europe) and lifestyle (lab strains, wild strains, sake strains, palm wine strains, and other fermentation strains). Furthermore they are genetically diverse, as evidenced by their lack of clustering within the *S. cerevisiae* phylogeny (Figure S1). This broad distribution across the species suggests that the AA112Δ allele is present at appreciable frequency in many populations of *S. cerevisiae.*

To further investigate this, we sequenced the *ERG28* promoter in EM93, the wild strain that accounts for ~88% of the BY genome [38-39]. Since EM93 is a diploid, we sequenced the promoter in the four spores from a single EM93 tetrad, in order to capture both alleles with no ambiguity. We found that the AA112Δ mutation was heterozygous within EM93, supporting our inference that it is commonly found in the wild. Together, these results suggest that the selection on *ERG28* in the BY lineage [8] was likely acting on standing variation, as opposed to a new mutation. Because EM93 is heterozygous, we can infer the selective sweep most likely occurred in the descendants of EM93, after its introduction to the laboratory.

To attempt a similar analysis for the seven other ERG genes involved in this adaptation (Figure 1A), we sequenced their promoters in the same four EM93 spores. Because we do not know the causal variants, we performed this analysis at the level of promoter haplotypes (sets of co-occurring alleles). We found that for all seven genes, the complete BY promoter haplotype was either homozygous (for two genes, *ERG25* and *ERG26*) or heterozygous (for five genes) in EM93, indicating that their *cis*-acting down-regulations were likely not due to new mutations occurring in the lab. Each of these BY haplotypes was also observed in between zero and six additional sequenced strains, indicating that some of the haplotypes are segregating at an appreciable frequency in *S. cerevisiae*. However the absence of a complete BY haplotype does not imply the absence of the causal BY variant, since most ERG promoter variants are not in perfect linkage disequilibrium with their neighboring variants. For example, although the AA112Δ variant was found in 12 strains (Figure S1), only five of these also had the T229C variant (and thus the complete BY promoter haplotype). This highlights the importance of identifying causal variants in order to study the evolutionary histories of specific adaptations.

**Discussion**

We have identified the causal mutation underlying a *cis*-regulatory adaptation that affects the ergosterol biosynthesis pathway in yeast, and characterized its molecular mechanism of action. The mutation, a 2 bp promoter deletion, reduces the expression of *ERG28* by ~1.3-fold. This effect is mediated by two TFs, Mot3 and Sok2, which bind immediately adjacent to the deletion; these TFs bind and regulate the wild-type RM *ERG28* promoter, but not the *ERG28* AA112Δ promoter.

Although it may seem surprising that a 2 bp deletion outside of TF binding sites can have such a strong effect on TF binding, it is consistent with previous work. First, most between-strain variation in the binding of the Ste12 TF in yeast cannot be linked to variation in any known TF motif, even when only considering those binding sites where occupancy was associated with nearby genetic markers [40]. Second, it was recently shown that changes in the positions of TF binding sites as small as 1-2 bp can result in substantial (>1.5-fold) effects on transcription [41]. Finally, minor changes in the copy number of very short tandem repeats in yeast promoters can also impact transcription [42].

It is also at first counterintuitive that decreased binding of a repressor (Mot3) could contribute to the down-regulation of *ERG28* by AA112Δ, in particular since the repressive effects of Mot3 appear to be stronger than the activation by Sok2 (Figure 3C). We hypothesize that the AA112Δ mutation may have altered the TF binding landscape upstream of *ERG28*, not only for Mot3 and Sok2, but possibly for other TFs as well. The deletion's effect on transcription would then be determined by this altered landscape.

In addition to the focus on *ERG28*, our results also further characterize the polygenic ERG pathway adaptation as a whole. We found that two genes not implicated in our previous analysis of microarray data [8], *ERG25* and *ERG27*, also show reduced expression from the BY allele (compared to RM). Moreover, our precise measurements of the *cis*-acting effect size for each ERG gene led us to an intriguing discovery: the five proteins that form the core of a complex at the ER membrane are also the five with the strongest *cis*-regulatory change. This pattern suggests an exquisite specificity of selection, in which the precise level of down-regulation is determined not only by position within the pathway, but also by membership in a specific protein complex.

While a handful of causal mutations underlying *cis*-regulatory adaptations in other model organisms have been previously reported [10-11], their molecular mechanisms are unknown. Compared to these, our knowledge of the *ERG28* AA112Δ mutation is now relatively detailed, though still incomplete; for example, how the deletion disrupts binding has not been established. A plausible explanation is that Sok2 and Mot3 may bind cooperatively to the *ERG28* promoter in wildtype RM; if this cooperativity is disrupted by the 2-bp deletion (which moves the binding sites closer together and changes their relative angles by ~70º), then neither factor would bind well to the AA112Δ promoter.

At the phenotypic level, we found that AA112Δ confers a condition-specific growth advantage in the presence of the antifungal drug amphotericin B. Because the AA112Δ mutation may also lead to a fitness advantage in other environments that were not tested, we cannot conclude whether amphotericin B is related to the specific selection pressure that gave rise to the ERG pathway adaptation in BY. However our results are quite consistent with previous observations that the down-regulation or deletion of ERG pathway genes confers resistance to amphotericin B in diverse clinical yeast isolates [18-21]. Thus our results may shed light on potential molecular mechanisms by which antifungal drug resistance can evolve.

## Materials and Methods

### *Strain construction*

We carried out all strain engineering in RM, as opposed to BY, because BY contains a very recent loss-of-function transposon insertion in the transcription factor *HAP1*, which alters the regulation of many ERG genes, including *ERG28*. Because this mutation was so recent (not even present in the very closely related lab strain W303 [8]), it must have happened after the *ERG28 cis*-regulatory adaptation, so the functional *HAP1* in RM should more accurately reflect the original effects of any *cis*-regulatory mutations.

*In vivo* site-directed mutagenesis, known as *delitto perfetto*, was performed as described [27]. Briefly, the pCORE-UH cassette, containing *K. lactis URA3* and *hyg*, was amplified using primers containing ~70 bp of homology to the RM ERG28 promoter (Table S1). This PCR product was transformed into RM, and correct incorporation into the *ERG28* promoter was verified by PCR. The site of incorporation was chosen in between the two candidate genetic variants, so that the same CORE cassette transformant could be used for engineering both mutations. The CORE cassette was then removed by separately transforming two PCR products from the BY *ERG28* promoter, containing the desired mutation (either AA112Δ or T229C) as well as enough flanking DNA sequence (identical between RM and BY) to allow specific targeting of the PCR product. Because the efficiency of *delitto perfetto* is maximized when transforming longer DNA molecules, as well as double-stranded DNA [20], transforming long PCR products from BY (as opposed to shorter, single-stranded synthetic oligonucleotides) is a useful modification. Counter-selection of the resulting transformants on 5-FOA allowed isolation of successfully engineered strains that had replaced the CORE cassette with the desired mutation, which were then sequence-verified.

The complete coding regions of *MOT3* and *SOK2* were replaced with the *hphMX6* antibiotic resistance gene via PCR-mediated gene disruption [43] in both RM and RM AA112Δ. Transformants were grown on hygromycin B, and verified by PCR. These two TFs were also HA-tagged at their C-termini via transformation of a PCR product including the HA tag, *hphMX6*, and flanking regions with 40 bp of homology to the targeted regions [43]. Transformants were grown on hygromycin B, and then verified by PCR and sequencing.

Table S2 lists all strains used in this work.

*Growth conditions*

With the exception of growth rate experiments (Figure 4), all strains were grown in standard YPD media at 30°C, and harvested in log-phase (OD600 ~1) for either RNA extraction or chromatin immunoprecipitation.

*RNA extraction and cDNA synthesis*

We extracted total RNA with the Epicentre Biotechnologies RNA Purification kit, which includes a DNase treatment to remove contaminating genomic DNA. RNA concentration was quantified with a NanoDrop2000 spectrophotometer. For cDNA synthesis, total RNA samples were diluted to a concentration of 500ng/µL. RNA was reverse transcribed into cDNA with SuperScript III RT (Invitrogen), following manufacturer protocols.

*Pyrosequencing*

Pyrosequencing was performed on a PyroMark Q24 (Qiagen), following manufacturer's protocols. Primers (Table S1) were designed to target individual SNPs in transcribed regions using the PyroMark Assay Design Software (Qiagen). Negative controls using no primers, or no cDNA template, were performed for each assay.

*Quantitative PCR*

cDNA was diluted 1:100 prior to qPCR. qPCR was performed on an Eco Real-Time PCR machine (Illumina) following manufacturer's protocols. To quantify changes in *ERG28* mRNA abundance, six control genes previously noted for their stability across conditions [44] were measured in each experiment: *ACT1, TDH3, ALG9, TAF10, TFC1,* and *UBC6*. All experiments were done in at least biological triplicate and technical duplicate. Experiments in Figure 3c were done in biological sextuplicate and technical quadruplicate. Data were analyzed using qBase Plus software (Biogazelle) [45].

*Chromatin immunoprecipitation*

Chromatin immunoprecipitation (ChIP) was performed essentially as described [46]. Briefly, wildtype cells and cells expressing either Mot3-HA or Sok2-HA were grown to mid-log phase in 100mL YPD. Cross-linking was performed by treating yeast with 1% formaldehyde for 15 minutes at 25°C. Chromatin was isolated from whole-cell extracts generated by spheroplasting and sheared by sonication. Immunoprecipitation was performed from 5µg chromatin using mouse monoclonal anti-HA (Invitrogen, clone 5B1D10) and immune complexes were captured with Ultralink Immobilized Protein A/G resin (Pierce). Protein-DNA complexes were eluted with 1% SDS/0.1 M $NaHCO_3$. Eluates were incubated at 65°C overnight to reverse cross-links and treated with proteinase K (Invitrogen) and RNAse A. DNA was phenol-chloroform extracted, ethanol-precipitated, and resuspended in water prior to qPCR.

ChIP DNA was amplified on an Eco Real-Time PCR machine (Illumina) following manufacturer's protocols. We quantified the abundance of the *ERG28* promoter region containing the Mot3 and Sok2 binding sites, as well as part of the *ACT1* coding region as a control to quantify the amount of DNA in each reaction. The concentration of *ERG28* promoter DNA was normalized against this control before comparing across genetic backgrounds (RM vs. RM AA112Δ) for each TF.

*Quantitative growth rate measurements*

To perform quantitative growth rate measurements (Figure 4), we grew strains in 96-well plates and measured OD600 at 15-minute intervals using an automated plate reader (Tecan) until cultures reached saturation. Data shown in Figure 4 are the mean $\log_2$ ratios of the maximum log-phase growth rates (estimated by Magellan software, Tecan) for 48 replicate growth curves of each strain. Growth conditions were SD media alone or 0.8 ug/ml amphotericin B in SD media, both at room temperature (22°C). P-values were calculated using a paired t-test, pairing wells in the same row on each plate. Other conditions tested in an initial screening phase were hyperosmotic stress (NaCl or menadione) and temperature stress (heat/freezing).

**Acknowledgements**

We thank D. Leifer and S. Bhutra for assistance with strain construction; S. Patel for assistance with pyrosequencing; and A. Ting for advice.

## Figure Legends

**Figure 1.** A polygenic gene expression adaptation in the ergosterol biosynthesis (ERG) pathway. **(A)** The final steps of the ERG pathway. Eight genes whose down-regulation contributes to a polygenic gene expression adaptation are colored red; the six previously implicated genes [8] are underlined. Erg28 is shown next to its strongest interaction partner, Erg27. **(B)** Allelic bias of ERG genes, as measured by pyrosequencing in the RM/BY hybrid. The allelic bias indicates the magnitude of *cis*-regulatory divergence between RM and BY for each gene. Red color indicates genes that are part of the polygenic adaptation. Asterisks indicate those that interact strongly with Erg28 [24], all of which have stronger allelic bias than those that do not.

**Figure 2.** Pinpointing the causal mutation affecting *ERG28 cis*-regulation. **(A)** Sequence divergence between RM and BY in the *ERG28* promoter region. No other differences exist for 590 bp upstream of the gene, or in the 5' UTR. **(B)** Genotypes at the two variable positions for RM, BY, and the two engineered strains. **(C)** The mRNA levels of *ERG28* in each of the two engineered strains compared to wildtype RM, assayed by qPCR. The causal mutation is expected to result in a ~1.30-fold difference, matching the allelic bias observed in the RM/BY hybrid (Figure 1B), whereas any non-causal mutation will not alter the RM expression level (~1-fold change). **(D)** Allelic expression bias in hybrids between each engineered strain and BY, assayed by pyrosequencing. Any non-causal mutation will not alter the 1.30-fold RM/BY allelic bias, whereas the causal mutation is expected to be expressed at the same level as the BY allele (~1-fold allelic bias).

**Figure 3.** Determining the molecular mechanism of the causal mutation. **(A)** Two predicted transcription factor (TF) binding sites flanking the deletion. **(B)** The expected fold-change in *ERG28* expression level when deleting TFs under different scenarios. Left: if a TF does not regulate *ERG28*, its deletion should have no effect on *ERG28* levels. Center: If a TF regulates *ERG28* and acts independently of the two-base deletion, then deleting the TF should result in some fold-change X, which will be observed in both the wildtype RM and RM AA112Δ backgrounds. Right: If a TF regulates the wildtype *ERG28* promoter, but the deletion abolishes this regulation, then the TF deletion may only affect *ERG28* mRNA levels in the wildtype

background (A fourth possible scenario, not shown, is where the TF only regulates *ERG28* in RM AA112Δ). **(C)** qPCR data showing changes in *ERG28* mRNA levels upon deleting either *SOK2* or *MOT3*. In both cases, a difference is observed in the wildtype background ($p = 7.5 \times 10^{-5}$ for *SOK2* and $5.0 \times 10^{-3}$ for *MOT3*), but not the RM AA112Δ background ($p = 0.28$ for *SOK2* and 0.67 for *MOT3*), consistent with the TF regulation being entirely abolished by the deletion. **(D)** Chromatin immunoprecipitation data showing the difference in binding for Sok2 and Mot3 to the *ERG28* promoter in wildtype RM / RM AA112Δ. In both cases a significant decrease in binding is observed in RM AA112Δ.

**Figure 4.** Fitness effect of the causal mutation. **(A)** In rich synthetic defined (SD) media, the RM AA112Δ strain and RM T229C strains show no significant difference from RM. **(B)** In the presence of the antifungal drug amphotericin B, the RM AA112Δ strain shows a growth rate advantage over RM, whereas the RM T229C strain shows no difference from RM. Bars represent the mean $\log_2$ ratios of maximum log-phase growth rates from 48 replicate cultures, +/- 1 S.E.

**Figure S1.** Phylogeny of *S. cerevisiae* (adapted from [37]), with strains containing the AA112Δ allele marked with an asterisk. Branch lengths are not to scale. “SGD” and “S288c” are nearly identical to BY, so are not counted among the 12 non-BY strains with AA112Δ.

**Table S1.** All PCR primers used in this work.

**Table S2.** All yeast strains used in this work.

A.

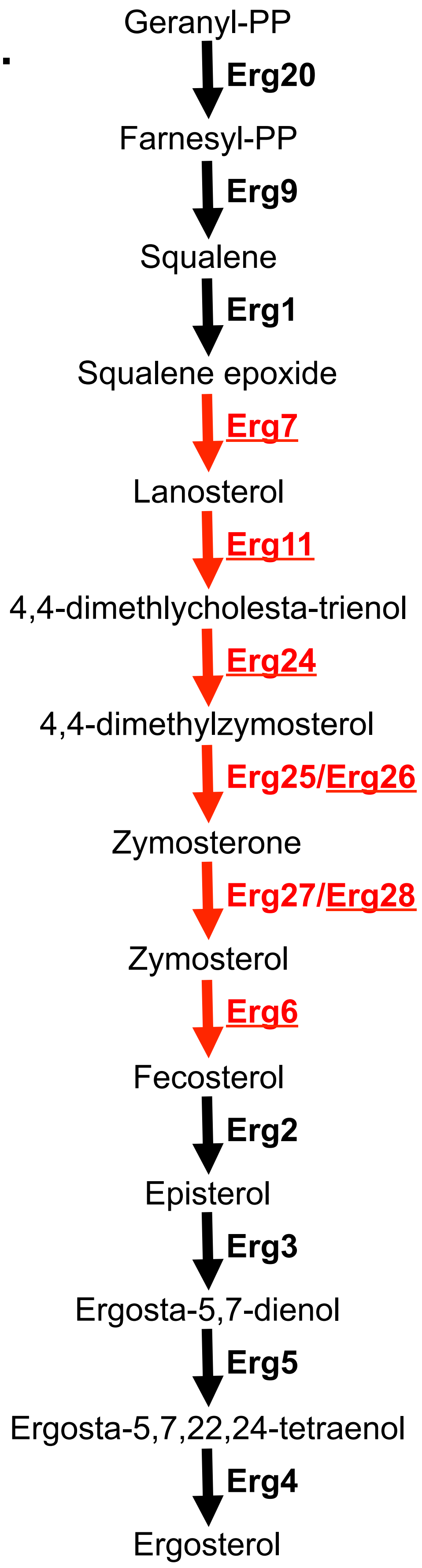

B.

| Gene | RM/BY allelic ratio |
|---|---|
| ***ERG7*** | **1.13** |
| ***ERG11**** | **1.96** |
| ***ERG24*** | **1.22** |
| ***ERG25**** | **1.47** |
| ***ERG27**** | **1.25** |
| ***ERG28**** | **1.30** |
| ***ERG6**** | **1.42** |
| ***ERG2*** | **0.97** |

***Interacts strongly with Erg28**

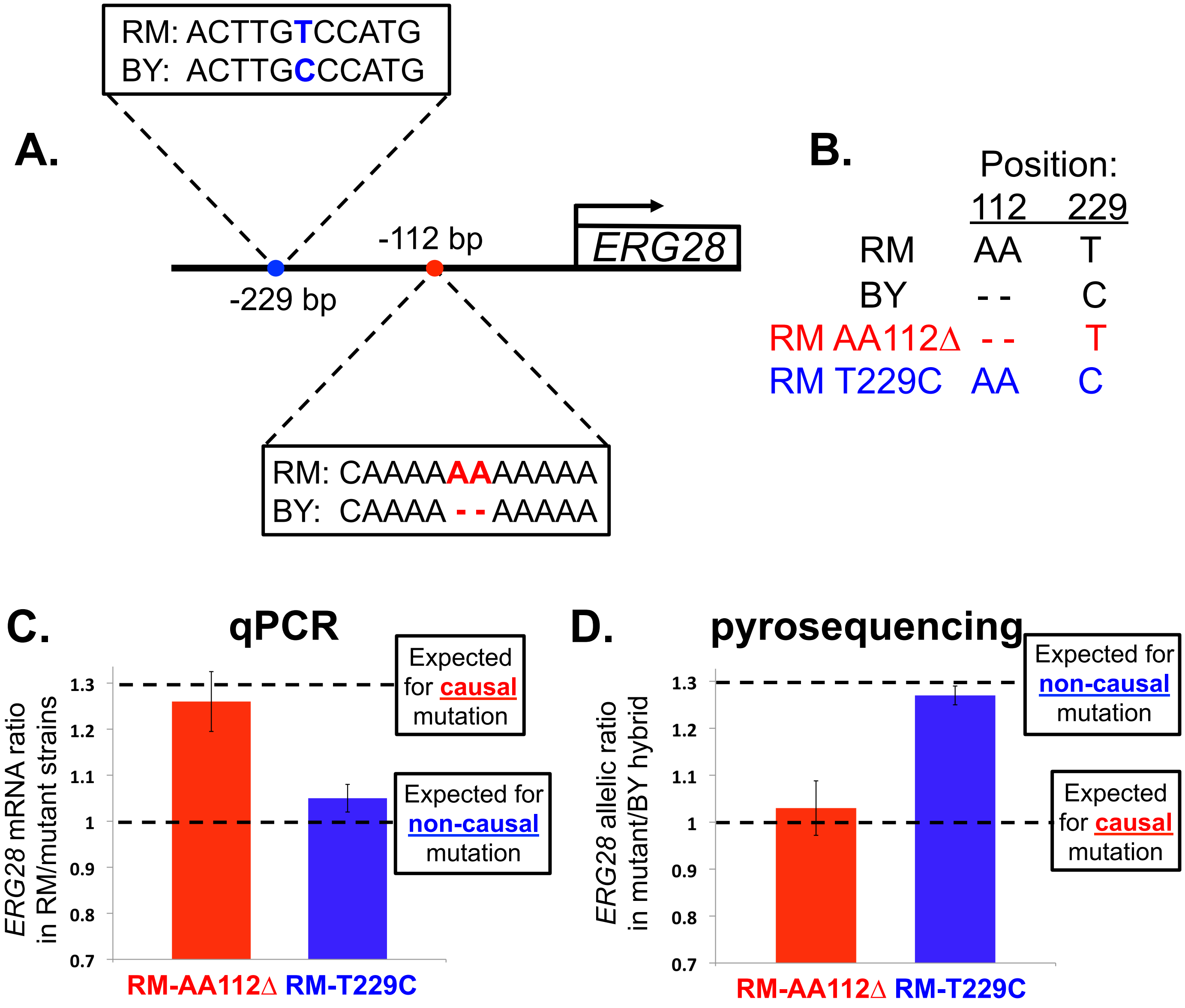

A.
RM: ACTTGTCCATG
BY: ACTTGCCCATG
-112 bp
ERG28
-229 bp
RM: CAAAAAAAAAAA
BY: CAAAA - - AAAAA
B.
Position:
112 229
RM AA T
BY - - C
RM AA112Δ - - T
RM T229C AA C
C.
qPCR
Expected for causal mutation
Expected for non-causal mutation
ERG28 mRNA ratio in RM/mutant strains
1.3
1.2
1.1
1
0.9
0.8
0.7
RM-AA112Δ RM-T229C
D.
pyrosequencing
Expected for non-causal mutation
Expected for causal mutation
ERG28 allelic ratio in mutant/BY hybrid
1.3
1.2
1.1
1
0.9
0.8
0.7
RM-AA112Δ RM-T229C

**A.**

Sok2 Mot3

RM: CGAA CAGGAAACA AAA**AA**AAA AAGGTAC GAT

BY: CGAA CAGGAAACA AAA - - AAA AAGGTAC GAT

**B.**

*ERG28* expr ratio in TF / *tf*Δ strain expected if:

| | TF does not regulate *ERG28* | TF regulation **independent** of deletion | TF regulation **dependent** on deletion |
|---|---|---|---|
| wild-type RM | 1 | X ≠ 1 | X ≠ 1 |
| RM AA112Δ | 1 | X ≠ 1 | 1 |

**C.**

*ERG28* expr ratio:

| | **SOK2 / *sok2*Δ** | **MOT3 / *mot3*Δ** |
|---|---|---|
| wild-type RM | 1.21 | 0.54 |
| RM AA112Δ | 1.01 | 1.02 |

**D.**

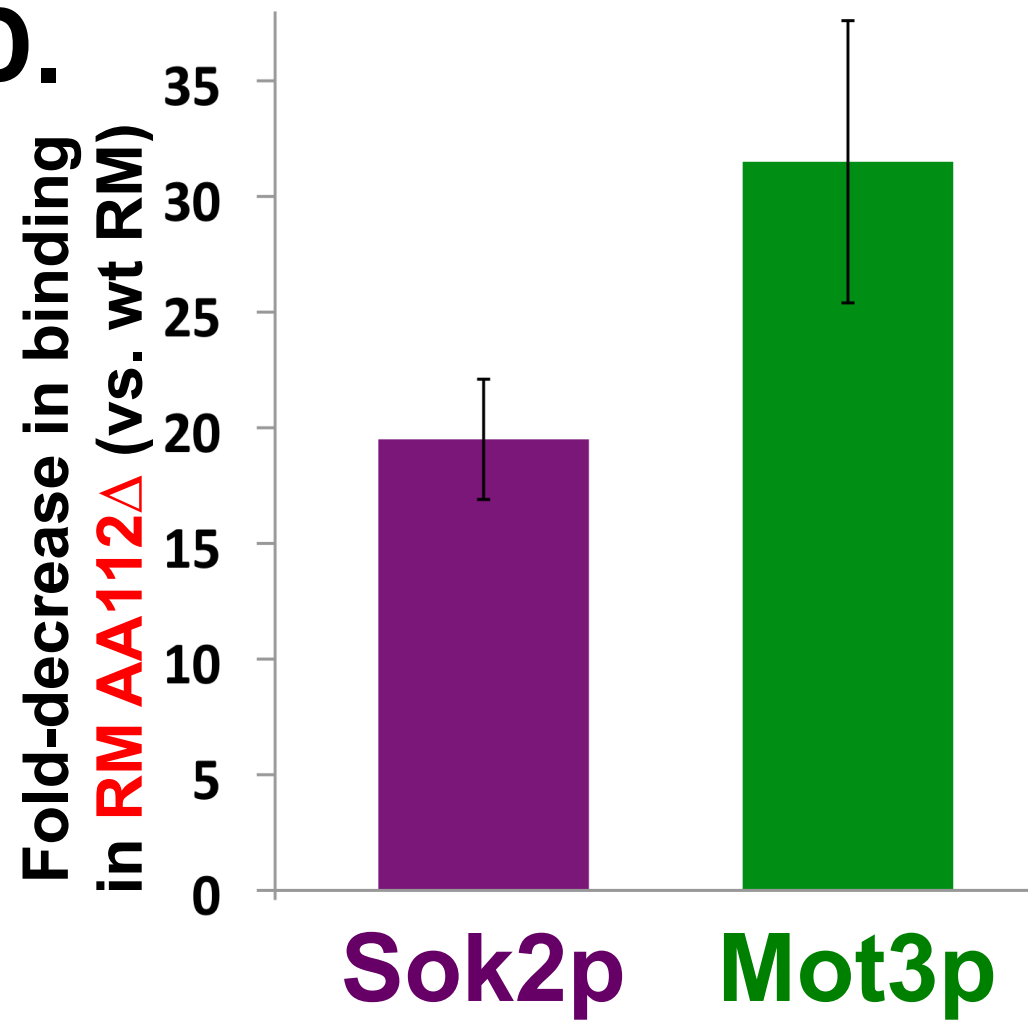

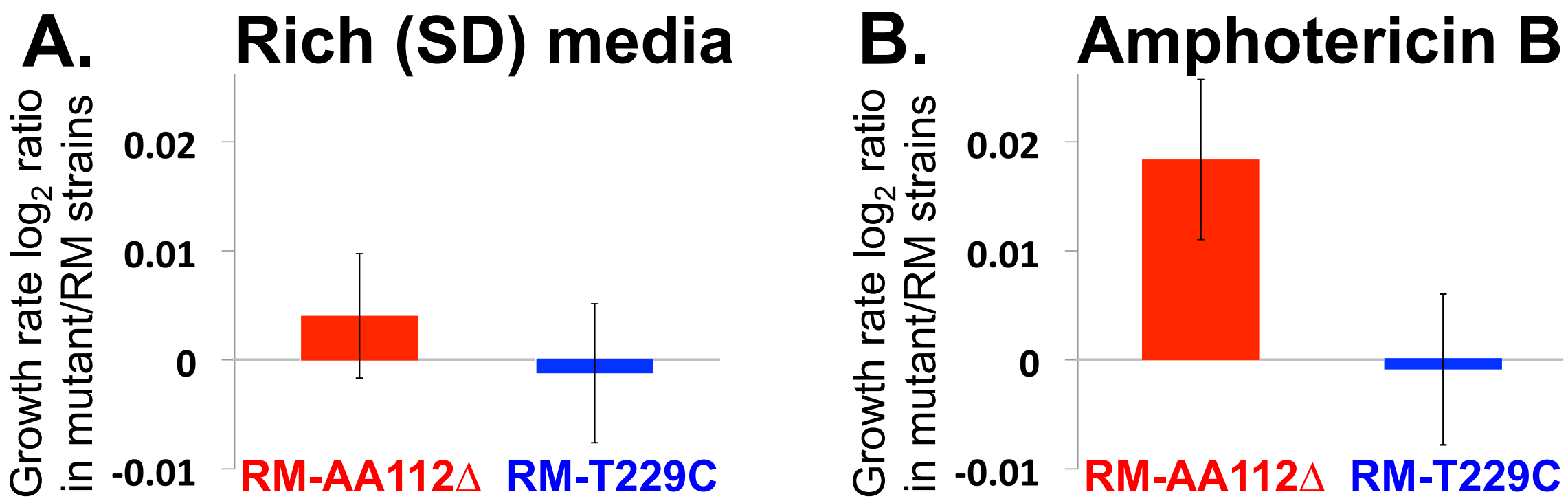